\documentclass[conference]{IEEEtran}
\IEEEoverridecommandlockouts
\usepackage{ifpdf}
\usepackage{cite}
\usepackage{amsmath,amssymb,amsfonts}
\usepackage{algorithmic}
\usepackage{graphicx, adjustbox}
\usepackage{array}
\usepackage{enumitem}
\usepackage{breqn}
\usepackage{ dsfont }
\usepackage{ upgreek }
\usepackage{textcomp}
\usepackage{multirow}
\usepackage{algorithm}
\usepackage{algorithmic}
\usepackage{caption}
\usepackage{pgfplots}
\usepackage{tkz-euclide}
\usepackage{tikz}
\usepackage{subfigure}

\begin{document}
\title{Blockchain as a Service for Multi-Access Edge Computing: A Deep Reinforcement Learning Approach
}
\author{\IEEEauthorblockN{ Dinh C. Nguyen$^1$, Pubudu N. Pathirana$^1$, Ming Ding$^2$, Aruna Seneviratne$^3$}
\IEEEauthorblockA{{$^1$School of Engineering, Deakin University, Waurn Ponds, VIC 3216, Australia } \\
\text{$^2$Data61, CSIRO, Australia  }\\
\text{$^3$School of Electrical Engineering and Telecommunications, UNSW, NSW, Australia  }
\\
\text{Email: \{cdnguyen, pubudu.pathirana\}@deakin.edu.au, ming.ding@data61.csiro.au, a.seneviratne@unsw.edu.au }
}
\thanks{*This work was supported by CSIRO Data61, Australia. }
}
\maketitle
\begin{abstract}
Recently, blockchain has gained momentum in the academic community thanks to its decentralization, immutability, transparency and security. As an emerging paradigm, Multi-access Edge Computing (MEC) has been widely used to provide computation and storage resources to mobile user equipments (UE) at the edge of the network for improving the performance of mobile applications. In this paper, we propose a novel blockchain-based MEC architecture where UEs can offload their computation tasks to the MEC servers. In particular, a blockchain network is deployed and hosted on the MEC platform as Blockchain as a Service (BaaS) that supports smart contract-based resource trading and transaction mining services for mobile task offloading. To enhance the performance of the blockchain-empowered MEC system, we propose a joint scheme of computation offloading and blockchain mining. Accordingly, an optimization problem is formulated to maximize edge service revenue and blockchain mining reward while minimizing the service computation latency with respect to constraints of user service demands and hash power resource. We then propose a novel Deep Reinforcement Learning (DRL) approach using a double deep Q-network (DQN) algorithm to solve the proposed problem. Numerical results demonstrate that the proposed scheme outperforms the other baseline methods in terms of better system utility with computational efficiency. Experiment results also verify that the trading contract design is efficient with low operation cost, showing the feasibility of the proposed scheme. 
\end{abstract}
\section{Introduction}

Recent years have witnessed the explosion of mobile technologies with the proliferation of user equipments (UEs) such as smartphones, tablets, wearable devices, etc, which have been driving the evolution of Internet of Things (IoT) [1]. UEs can be employed to run IoT applications such as smart home, smart healthcare, smart city with high flexibility and efficiency [2]. However, due to the rapid growth of mobile data traffic in mobile networks, i.e. 5G and beyond, executing extensive IoT applications merely on overloaded UEs is incapable of providing satisfactory Quality of Services (QoS) to users. Fortunately, the recent advances of mobile edge technologies enable UEs to offload their computation-extensive tasks (i.e. data and programming codes) to Mobile Edge Computing servers with a close proximity to UEs. Mobile Edge Computing was conceptualized by the European Telecommunications Standards Institute (ETSI) Industry Specification Group (ISG) as a means of extending intelligence to the edge of the network along with better computing and storage capabilities. Since 2017, the ETSI industry group renamed it to Multi-Access Edge Computing (MEC), because the MEC technology brought benefits beyond mobile devices and to Wi-Fi/fixed access technologies [3]. The underlying concept of MEC is to extend cloud computing capabilities to the edge of cellular networks. This solution potentially reduces computation pressure on devices and provides ultra-low latency mobile services to satisfy the ever-increasing computation demands of modern UEs. As in previous works [4], [5], [6], many approaches using MEC have been proposed to improve IoT offloading performances from various aspects, such as computation efficiency, resource allocation and user selection. 

Although MEC plays an important role in IoT offloading, some critical challenges remain unsolved. First, it is insecure for UEs to perform large-scale offloading to MEC servers in untrusted and non-transparent environments. Second, MEC servers may not be willing to participate as computation service suppliers for UEs's offloading requests due to unfair resource allocation [5], which potentially degrades the availability of edge computing services. Thirdly, in the traditional computation offloading procedure, there is the need for a third-party intermediary to take the role of controlling resource trading (i.e. offloading payment). However, this architecture remains single-point failure issues when the intermediary is out of service as well as privacy bottlenecks due to curious third parties [7]. 

Blockchain that is a decentralized public ledger with high security capabilities has emerged as a promising solution to overcome these challenges [7]. The key concept of blockchain is based on a peer-to-peer network architecture in which transaction information is distributed among multiple nodes and not controlled by any single centralized entity. Specially, the blockchain network can be deployed and hosted on a MEC platform as Blockchain as a Service (BaaS) which enables blockchain-related services such as user access monitoring, transaction mining, and ledger network management [7]. BaaS with its decentralized and trustworthy capabilities has been considered for MEC systems for security guarantees such as offloading traceability [8], secure resource trading [9] and smart contract-based access control [10]. 

\subsection{Related works}
Many research efforts are directed to the design of MEC offloading with blockchain. The authors in [8] considered a blockchain-enabled computation offloading framework in multi-user MEC where task offloading time, energy consumption and maintaining load balance are optimized with respect to data integrity guarantees using blockchain. The work in [9] focused on solving resource trading issues by taking advantage of a smart contract which enables intelligent resource management and high security for the involved offloading systems. To solve the problem of resource scarcity during the offloading on MEC, the authors in [11] proposed a joint framework of coin-loaning and computation-offloading for blockchain-empowered MEC. The key objective is to minimize users' costs (i.e. energy consumption and network latency) by leveraging a game theory approach. Further, to achieve better resource management for multiple access mobile edge computing on blockchain, an economic approach was proposed in [5] to optimize the blockchain mining rewards of mobile miners through a Stackelberg game model. 

In comparison to such approaches, Reinforcement Learning (RL) [12] emerges as a strong alternative to solve offloading problems. However, the technical characteristics of multi-user MEC can make it challenging to apply RL, as the dimension of state and action space increases exponentially with the increasing number of users. Deep Reinforcement Learning (DRL) approaches such as deep Q-network (DQN) [13] have been employed to solve the above complicated problems. Our works [14, 15] implemented mobile blockchain-edge offloading frameworks using DRL with the main objective of minimizing computation cost and enhancing user privacy levels. The works [16, 17] employed DRL to build offloading algorithms for mining tasks and data processing tasks in blockchain. 

Although the literature works have addressed blockchain-based MEC offloading issues, there remains some critical issues that need to be solved. First, in most current studies [8], [16-17], MEC servers are always allocated randomly hash power resource for block mining regardless of their computation contributions, which leads to the unfairness among MEC servers. Further, considering the complex MEC system with multiple MEC servers and UEs (i.e. in 5G scenarios), the state and action space would be very large, which poses challenges for effectively solving the optimization problem. Besides, the existing approaches require static operating models [9], [11-12] which need to be updated over time. These schemes do not scale well when the dimension of the MEC system increases with the increase the number of MEC servers and UEs associated with a larger number of configuration parameters. Final, the current DRL-based schemes [14-17] have not considered the joint design of computation offloading and blockchain mining, which has significant impacts on the performance of blockchain-empowered multi-user MEC systems. 

\subsection{Contributions} 
Different from the literature works, in this paper, we propose a new blockchain-based mutative MEC system for supporting computation offloading by using BaaS services, including shared transaction ledgers, smart contracts, and transaction consensus. UEs can submit their offloading requests to the blockchain by establishing a transaction which is transmitted to MEC servers via blockchain ledgers for traceability and security. Moreover, smart contracts are leveraged to enable reliable resource trading (i.e. payment) between UEs and MEC servers without the need for any intermediary. MEC servers use their resources to execute the offloaded tasks. Given the resource constraints of mobile devices, we introduce an edge mining solution which enables MEC servers to participate in the consensus process for extra profits through mining rewards. In particular, we propose a reputation mechanism that evaluates the computation contribution of MEC servers to adjust adaptively the resource allocation in the MEC network. This solution would accelerate the mining process and encourage more MEC servers to participate in the offloading process for improving network robustness. Further, we also apply the Software Defined Networking technology to provide logical control to our MEC architecture. The DRL unit is also deployed on the SDN controller that provides a holistic overview of the network states to better manage offloading and mining services. We consider that the system performance is constrained by users' service demands and resource (i.e. hash power) of MEC servers. Then, the user selection and  resource allocation are formulated as a joint optimization problem. We then propose a novel deep reinforcement learning (DRL)
scheme using a double DQN algorithm to solve the proposed problem. The simulation results verify the high performance of our proposed algorithm, compared to other baseline approaches. 

The main contributions of this paper can be summarized as follows.
\begin{enumerate}
	\item We propose a new multi-access edge computing (MEC) architecture with blockchain for multi-user networks. To enhance the performance of the blockchain-empowered MEC system, we propose a joint scheme of computation offloading and blockchain mining. Particularly, we leverage a smart contract to support transparent resource trading for computation offloading, while a reputation mechanism is designed to accelerate the mining process. 
	\item	We formulate an optimization problem that aims to maximize edge service revenue and blockchain mining reward while minimizing the service computation latency with respect to the constraints of user service demands and hash power resource. 
	\item Then, we propose a novel Deep Reinforcement Learning approach using a double DQN algorithm to solve the proposed problem. Numerical simulation results demonstrate that the proposed scheme yields the best system utility performance with minimal computation complexity, compared with the other baselines. Experiment results also verify the efficiency of the smart contract design with low operation costs.
\end{enumerate}
The remainder of the paper is organized as follows. Section II introduces the BaaS service used in our work and describes the system concept. We then present the system formulation in terms of computation offloading and blockchain mining in Section III. We detail the proposed DRL scheme and present a double DQN algorithm to solve the proposed problem in Section IV. Section V presents simulation results and discussions on the obtained results, and Section VI concludes the paper. 

\begin{figure}
	\centering
	\includegraphics[ height=5.6cm, width=5.3cm]{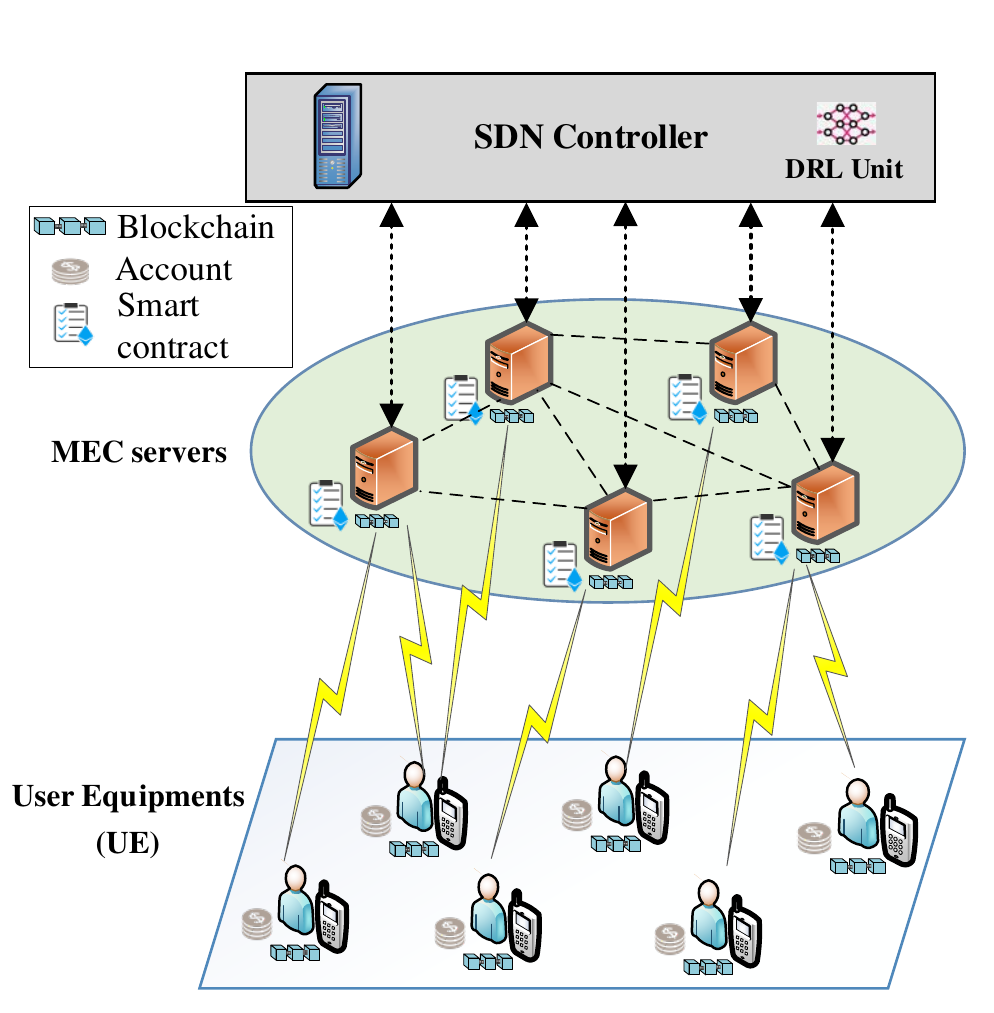} 
	\caption{The proposed blockchain-empowered MEC system. }
	\vspace{-0.25in}
\end{figure}

\section{	System description}
In this section, we explain the BaaS terminology used in our work and then introduce the system concept.
\subsection{	Blockchain as a Service for MEC}
In our paper, a blockchain network is deployed and hosted on a MEC platform as Blockchain as a Service (BaaS). In particular, BaaS can offer a number of blockchain-enabled services to support mobile applications.
\begin{itemize}
	\item \textit{Shared ledger:} It represents a database that is shared and distributed among edge network participants (i.e. UEs and MEC servers). The shared ledger records transactions that contain request information of UEs (i.e. user ID) and offloading information among UEs and MEC servers (i.e. service demand, data size). Importantly, the shared ledger interconnects the MEC servers in a secure manner, and each MEC server has an equal role in managing the MEC network without the need of external authorities according to the blockchain concept. Thus, the use of shared transaction ledger via blockchain eliminates single-point failure risks faced by traditional MEC systems [4], [6].
	\item \textit{Shared smart contract: }Blockchain also offers smart contract services for computation offloading. In specific, we use a smart contract as an intelligent middle layer to implement reliable resource trading, and UEs can perform payment to the edger service provider (ESP) through their blockchain account. An advantage of a smart contract is the transparency that means both UEs and ESPs can monitor and trace the payment process via shared transactions on the blockchain.
	\item \textit{	Consensus: }Our BaaS model also provides a consensus service on the MEC platform that is highly necessary for our MEC system in improving blockchain consistency and ensuring high network security. By using a reputation score mechanism based on data computation contributions, the MEC servers can compete together to join the block mining process for extra profits (i.e. reward in coins). Details of the concept will be explained in the following section.
\end{itemize}
\begin{figure}
	\centering
	\includegraphics[ height=2.3cm, width=8.5cm]{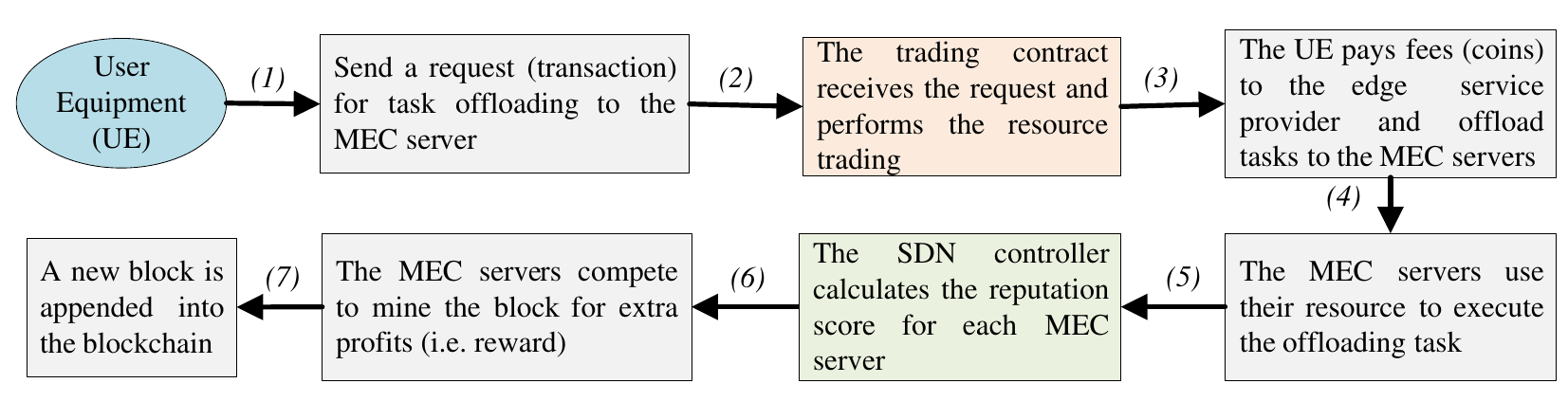} 
	\caption{The system concept.}
	\vspace{-0.25in}
\end{figure}
\subsection{System concept}
We propose a mobile cloud architecture as shown in Fig. 1. Without loss of generality, we assume that there are $M$ MEC servers, denoted as $\mathcal{M} = \{1,2,..., M\}$. They provide computation services for UEs and also perform blockchain consensus for obtaining rewards. We consider a set of UEs denoted as $\mathcal{U} = \{1,2,..., U\}$, and each UE can offload its computation tasks to the MEC servers for execution. Each MD has a blockchain account to join the blockchain network and perform trading which is enabled by a smart contract called \textit{trading contract}. We elaborate on the detailed operations of the proposed scheme in Fig. 2 and explained as follows.

\textit{(1)	System initialization:} Each UE $u$ needs to register with the blockchain network to become a legitimate participant with a uniquely identified address $Id_u$. By creating a blockchain account, which includes a private key $SK_u$ to sign the transaction and public key $PK_u$ for user identification [13], the UE $u$ can join the blockchain network. To realize the transactions of edge offloading, the UE $u$ also needs to get a wallet address $W_u$. As a result, the transaction detail of each UE $u$ can be expressed by a variable tuple $(Id_u, SK_u, PK_u, W_u)$, which is used when performing an offloading request to the MEC servers.

\textit{(2) Offloading to the MEC servers:} This process includes three phases: resource trading, edge computation and calculation of reputation score. First, to be served by the MEC servers, the UE $u$ needs to send a transaction containing an offloading request to the MEC servers. This is done by triggering the trading contract, which acts as a middle layer to verify the transaction (checking User ID, account balance) (see Steps 1,2 in Fig. 2). Details of the operation of our smart contract can be found in our recent work [10]. Then, the contract performs resource trading that means the user pays a certain fee (in coins) for the edge service provider (ESP) based on the service demand (Step 2 in Fig. 2). After successful payment, the UE now can offload its tasks to the MEC servers for execution. Then, the SDN controller calculates the reputation score of each MEC server according to the service latency (see Steps 3,4,5 in Fig. 2).  

\textit{(2)  Block mining: } Based on the reputation score, the SDN controller will allocate an appropriate hash power resource for each MEC server to perform mining. The MEC server with a higher reputation value will be allocated more hash resources for mining. Then, the new block is mined and appended into the blockchain (see Steps 6,7 in Fig. 2).  

\section{Problem formulation}

We assume that each UE $u$ has a computation task to be executed, denoted as $R_u = (d_u, \xi_u, \tau_u)$ where $d_u$ is the size of a task (in bits), $\xi_u$ is the number of CPU cycles of the task $R_u$ (service demand), and $\tau_u$ is the desired execution time to complete the task computation. Such information is known to the UE before the task is computed [11]. We denote $\mu$ as the price unit for computing services and this is charged by the ESP during resource trading. We consider realistic IoT scenarios such as healthcare [10] where complex data tasks (i.e. image extraction or video decoding) should be offloaded to MEC servers for efficient computation. Here we focus on formulation of the edge offloading and block mining for our blockchain-based MEC system.

\subsection{Offloading to MEC servers}
The offloading process is divided into three phases, namely resource trading, edge computation and score reputation evaluation.   

\textit{Resource trading cost}: To perform data offloading to the MEC servers, the UE $u$ needs to pay the service cost of executing its task to the ESP during the resource trading via the trading contract. This is also regarded as the revenue of the MEC system from computation services. Based on the service demand $\xi_u$ and unit price $\mu$, the service cost charged by the ESP to serve the UE $u$ is calculated as
\begin{equation}
\phi_u = \mu\xi_u.
\end{equation} 

\textit{Edge computation}: 
We model the edge computation cost for executing data tasks from UEs. Note that the edge execution happens only when the resource trading is completed. Considering that latency is far more important than the energy metric in analyzing the edge computation [5, 14], then we focus on analyzing the task latency metric in our MEC formulation. We denote  $f^m_{u}$ (in CPU cycles/s) as the MEC server $m$'s computing resource allocated to serve the UE $u$, the time consumed to execute the computation task of a UE $u$ by the MEC $m$ can be formulated as 
\begin{equation}
T^m_u = x^m_u\frac{d_u\xi_u}{f^m_u}, 
\end{equation}
where $x^m_u \in \{0,1\} $ determines that the MEC server $m$ serves user $u$ or not. Here, $x^m_u = 1$ means that the MEC server $m$ accepts to serve the user $u$, $x^m_u = 0$ otherwise. Note that at each certain time, each MEC server only serves a UE, thus $\sum_{u=1}^{U} x^m_u = 1, \forall m \in \mathcal{M}$. 

\textit{Reputation Evaluation Mechanism: }
The SDN controller is responsible to aggregate all computation information of MEC servers to evaluate the reputation score which determines that which MEC server should be allocated more network resource (i.e. hash power) according to the service latency. In specific, an edge reputation score ($\Upsilon^m$) is defined for completing a computation task by a MEC server $m$, which is expressed as $\Upsilon^m = t^m_{des}/t^m_{act}$ where $t^{des}$ and $t^{act}$ are desired service latency and actual service latency, respectively. The MEC server with a higher reputation score should be prioritized to be allocated more resources (i.e. hash power) for joining the mining process. Mathematically, the reputation score ($\Upsilon^m$) of the MEC server $m$ can be calculated as
\begin{equation}
\Upsilon^m = \frac{t^m_{des}}{t^m_{act}} =\frac{\sum_{u=1}^{U}\tau_u}{\sum_{u=1}^{U} T^m_u}.
\end{equation}

Our reputation mechanism can help overcome the limitation of the current works [11], [12], [16] in terms of mining speed. Indeed, in such works, hash power is usually allocated randomly to miners regardless of their contributions to the computation offloading process. This can result in a case that a certain miner spends more resource for task execution but receives less hash power, which prevents it from participating in the consensus process in the future. Consequently, less MEC servers join the mining that adversely affects the blockchain-based MEC system. By introducing a reputation mechanism, we will encourage more MEC servers to solve the blockchain consensus puzzle, which would accelerate the mining process and ensure the robustness of the MEC system. In the next section, we will analyze the relationship between the reputation score and hash power allocation.
\subsection{Block mining}
As part of the blockchain network, MEC servers can act as blockchain miners to join the consensus process in order to perform mining tasks for extra profits. In the blockchain network, MEC servers (miners) compete against each other to become the first one to solve the mining puzzle. Each MEC server $m$ has a hash power, denoted as $p_m$ (Hash/sec). In our design, the hash power of each MEC server is allocated by the ESP provider according to its reputation score. In other words, the MEC server with higher reputation score should be allocated more hash power for its task execution efforts. This mechanism is extremely important to motivate more MEC servers to participate in the consensus process. In return, they also receives benefits (i.e. mining reward) from joining the consensus process. Thus, our solution can significantly enhance the robustness of the edge blockchain system and effectively solve the issue of a lack of miners for maintaining the blockchain network.  

 Then, the relative hash power of the MEC server $m$ to the blockchain network can be defined as 
\begin{equation}
\upmu_m = \dfrac{p_m}{H}
\end{equation}
where $H$ represents the total hash power of the blockchain network which can be estimated through the compact status reports from miners [18]. It can be seen that when $\upmu_m$ increases, the probability to achieve successful consensus increases, which will increase the mining reward for the MEC server $m$, accordingly.

In general, in order to mine successfully a block, two steps are needed including the mining step and the propagation step. In the mining step, the probability that the MEC server $m$ mines the block is proportional to its relative hash power $\upmu_m$. The mined block is then propagated to the blockchain network. However, there is a possibility that this MEC server $m$ propagates the mined block slower with other MEC servers in the propagation step, which makes such a block likely to be discarded from the blockchain. This issue is called as orphaning [12], and this miner does not receive a reward for its mining. According to [12], the orphaning probability is approximated as $\mathcal{P}_{orphan} = 1- e^{-\upeta\upphi(b_m)}$, where $\upeta$ is a constant mean value and set $\upeta =1/600(sec)$ [14]. Further, $b_m$ denotes the number of transactions contained in the block that is mined by the miner $m$, and $\upphi(b_m)$ represents the block propagation time.

Clearly, with a larger block size $s_m$, the propagation time required to reach a consensus of a block will be larger. As a result, the probability of successful mining by miner $m$ can be expressed as 
\begin{equation}
\mathcal{P}= \upmu_m(1- \mathcal{P}_{orphan}) = \upmu_me^{-\upeta\upphi(s_m)}. 
\end{equation}

We denote $\mathcal{R}$ as the reward of the first miner which achieves consensus, then the expected mining reward of the MEC server $m$ can be calculated as 
\begin{equation}
	R_m = \mathcal{R}\mathcal{P} = \mathcal{R}\upmu_ne^{-\upeta\upphi(s_n)}. 
\end{equation}

From our analysis, we can see that the performance of our blockchain-based MEC system can be improved by maximizing the system revenue $ \phi_u$ and mining reward $ R_m$, while minimizing the edge computation cost (i.e. execution latency $ T^m_u$). To achieve this goal, the decision-making process can be taken into consideration. In specific, the user with higher service demands should be selected (in the offloading process), and the MEC server with a higher reputation score should be allocated more computation resources (i.e. hash power). Accordingly, the user selection and resource allocation should be jointly considered. Motivated by this, in the following section, we adopt a Deep Reinforcement Learning approach to solve the formulated problem. 
\section{	Deep Reinforcement Learning Approach}
In this section, we first introduce a reinforcement learning method and propose a new DRL approach for our formulated problem.
\subsection{Reinforcement Learning formulation}
In our blockchain-empowered MEC system, the DRL unit on the SDN controller acts as the agent which interacts closely with the multi-user MEC environment to find an action $a$ for a state $s$ using a policy $\pi$ as shown in Fig. 3. This policy is defined as a mapping from the action to the state, i.e., $\pi(s)=a$. The main goal of the agent is to find an optimal policy $\pi$, aiming to maximize the total amount of award $r$ over the long run. To implement the RL-based algorithms, we first define the specific state, action and reward for the proposed model.
\begin{figure}
	\centering
	\includegraphics[ height=7cm, width=8cm]{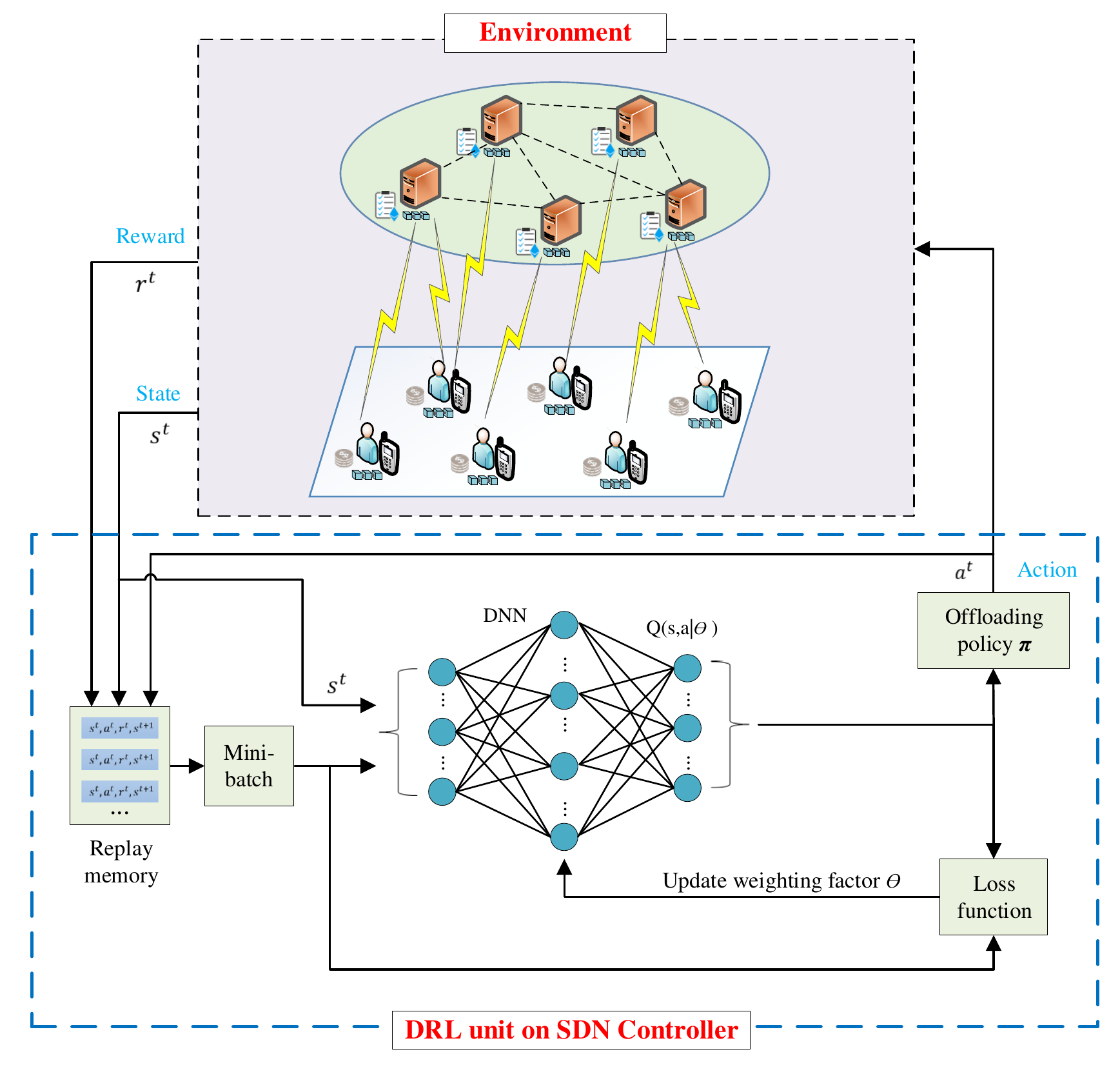}
	\caption{ The detailed DRL framework for blockchain-based MEC system. }
	\vspace{-0.1in}
\end{figure}

\subsubsection{State Space} The learning agent interacts with the environment to collect necessary information, including service demands of all UEs ($\xi_u, \forall u \in \mathcal{U}$), reputation scores of all MEC servers ($\Upsilon^m, \forall m \in \mathcal{M}$), which are expressed by matrix $S_U$ and $S_M$, respectively. In specific,  matrix $S_U$ can be defined as 
\begin{equation}
S_U = [s_1, s_2,..., s_u,...,s_U ],
\end{equation}
where $s_u$ presents the service demand of the UE $n$. Meanwhile, matrix $S_M$ can be specified as 
\begin{equation}
S_M = [s_1, s_2,..., s_m,...,s_M ],
\end{equation}
where $s_m$ is the reputation score of the MEC server $m$. Thus, the state space can be defined as 
\begin{equation}
s = [S_U, S_M].
\end{equation}

\subsubsection{Action Space} The learning agent needs to determine which user equipment is served by MEC servers (user selection) and how much mining computation resource (hash power) is allocated to MEC servers. As a result, the action space can be defined as
\begin{equation}
a = 
\begin{bmatrix}
x^m_1 & x^m_2 &...& x^m_u&...& x^m_U\\
p^1 & p^2 &...&p^m&...&p^M
\end{bmatrix},
\end{equation}
where $x^m_u \in \{0,1\} $ determines that the MEC server $m$ serves user $u$ or not. Here, $x^m_u = 1$ means that the MEC server $m$ accepts to serve the user $u$, $x^m_u = 0$ otherwise. Moreover, $p^m$ is the hash power allocated to the MEC server $m$ ($p^m < H$). 
\subsubsection{Reward Function}
The objective of the RL agent is to find an optimal policy $\pi^*$ which makes a decision $a$ at each state $s$ to obtain a certain reward that needs to reflect the objective of our DRL algorithm. This is to maximize the mining reward $R_m$ and service revenue $\phi_u$ while minimizing the computation cost $T^m_u$ for the long run. Accordingly, we define the expected system utility as the reward that the DRL agent receives from the environment as

\begin{equation}
r = \mathds{E}_{\pi,s}\sum_{t=1}^{T}\sum_{m=1}^{M}(R_m+ \phi_u-T^m_u), \forall u \in \mathcal{U},
\vspace{-0.06in}
\end{equation}
where $T$ is the time horizon of the optimization problem.
\subsection{	Proposed DRL-based algorithm}
In this section, we present the design of the proposed DRL algorithm for our MEC scheme. 
\subsubsection{Basics of DRL}
To address the formulated MDP, RL-based approaches can be used to obtain the sequence of actions to maximize the long-term reward in (6). In RL, a learning agent performs trial and error processes via jumping state by state in the defined MDP by taking some actions to obtain certain rewards. By continuously interacting with the MDP, the agent can accumulate much experience over time for building an optimal policy. Specifically, the state-action function can be updated using the experience tuple of an agent $(s^t,a^t, r^t, s^{t+1})$ at each time step $t$ as $Q(s^t,a^t) \leftarrow Q(s^t,a^t) + \alpha \sigma^t $
which is called as Q-learning algorithm [8]. Here,  $\sigma^t = r(s^t,a^t) + \gamma*maxQ(s^{t+1},a^{t+1}) - Q(s^t,a^t)$ is the TD error which will be zero for the optimal Q-value, $\alpha$ is the learning rate, and $\gamma$ is the discount factor between (0,1). 

It is noted that such experiences are stored in a two-dimensional action-state table, which may become infeasible to solve problems with large state-action space. In our scenario, for example, assuming the time horizon is 200 time slots and 10 mobile users (MDs), our scheme has as many as $10*200^5$ states and $10*2^2$ actions associated with each state, which is extremely large. Motivated by this, function approximators such as a deep neural network (DNN) has been proposed and proven success in recent works [13, 14]. In DRL algorithms, the experience replay solution is employed in the training phase with buffer $\mathcal{B}$ which stores experiences $e^t= (s^t,a^t,r^t,s^{t+1})$ at each time step $t$. A random mini-batch of transitions $(s^j,a^j,r^j,s^{j+1})$ from the replay memory is also selected to train the Q-network. Here the Q-network is trained by iteratively updating the weight $\theta$ to minimize the loss function, which is written as
\begin{equation} \footnotesize
L(\theta) = \mathds{E}  [r^j +\gamma minQ(s^{j+1},a^{j+1}|\theta')-Q(s^j,a^j|\theta^j))^2].
\vspace{-0.06in}
\end{equation}

However, the fact that the traditional DRL algorithm requires exhaustive search via all actions and states, it does not suit the high-dimensional problems like our offloading problem. Furthermore, during the computation of Q function in some MDP problems, it is unnecessary to estimate action and state values at the same time, thus we can estimate separately the action and state value functions. 
\subsubsection{Proposed double DQN algorithm}
In conventional DQN, we use the same samples from the replay memory to both specify which action is the best and estimate this action value, which leads to the large over-estimation of action values. To solve this problem, Double DQN proposes the use of two value functions with two sets of weights. One is utilized to determine the action, while the other is used to evaluate its reward. The double Q functions select and evaluate action values by the new loss function as follows

\begin{equation}
L_{dou}(\theta) = \mathds{E}  [y^j_{dou}-Q(s^j,a^j|\theta^j))^2],
\end{equation}
where $y^j_{dou}$ is specified as
\begin{equation}
y^j_{dou}= (r^j+\gamma.Q(s^{j+1},min_{a^{j+1}}Q(s^{j+1},a^{j+1}|\theta),\theta'). 
\end{equation}
 
It is noting that the action choice is still based on the weight $\theta$, while the evaluation of the selected action relies on the weight value $\theta'$. This technique reduces over-estimation problem and improves the overall performance of the learned model, compared with the model that uses traditional Q-learning. The details of the proposed algorithm are shown in Algorithm 1.

\begin{algorithm}
	\caption{The proposed DRL algorithm with Double DQN}
	\begin{algorithmic}[1]
		\STATE \textbf{Initialization:}
		\STATE Set replay memory $\mathcal{D}$ with capacity $N$
		\STATE Initialize the deep Q network  $Q(s,a)$ with random weight $\theta$ and $\theta'$, initialize the exploration probability $\epsilon \in (0,1)$
		\FOR{episode = 1,..., \textit{M}}
		\STATE Initialize the state sequence $s^0$
		\FOR{$t = 1,2,...$}
		\STATE \textit{/$***$ Plan the offloading and mining$***$/}
		\STATE Estimate the system state profile $s^t = \{S_U,S_M\} $
		\STATE Select a random action $a^t$ with probability $\epsilon$, otherwise $a^t = argminQ(s^t,a,\theta)$
		\STATE Perform the offloading process of trading, computation and mine the block
		\STATE Observe and evaluate the system utility $r^t$ based on (11) and  next state $s^{t+1}$
		\STATE \textit{/$***$ Update $***$/}
		\STATE Store the experience ($s^t,a^t,r^t,s^{t+1}$) into the memory $\mathcal{D}$
		\STATE Sample random mini-batch of state transitions ($s^j,a^j,r^j,s^{j+1}$) from  $\mathcal{D}$
		\STATE Calculate the target Q-value by ($y^j_{dou} = r^j+\gamma.Q(s^{j+1},min_{a^{j+1}}Q(s^{j+1},a^{j+1}|\theta);\theta')$
		\STATE Perform a gradient descent step with the weight $\theta^j$ on $ (y^j_{dou}-Q(s^j,a^j|\theta^j))^2$
		as the loss function
		\STATE Train the deep Q-network with updated $\theta$ and $\theta'$
		\ENDFOR 
		\ENDFOR 
	\end{algorithmic}
\end{algorithm}

\section{Performance Evaluation}
In this section, we evaluate the performance of the proposed algorithm through numerical simulations. 
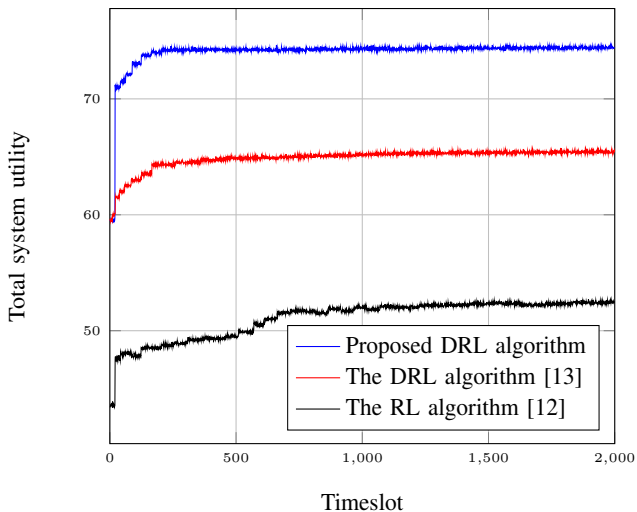
\begin{figure}
	\centering
	\setlength\abovecaptionskip{-0.05\baselineskip}
	\resizebox {8.8cm} {6.8cm} {
		\begin{tikzpicture}
		\tikzstyle{every node}=[font=\small]
		\begin{axis}[legend pos=south east, legend cell align={left},grid=major, xmax=2000,
		xmin=0,
		xlabel=Timeslot, ylabel = Total system utility,
		tick label style={font=\tiny} ]
		\addplot[blue, mark=none] table[x=t, y=output] {convergence_DRL.dat};
		\addplot[red,mark=none] table[x=t, y=output1] {convergence_DRL.dat};
		\addplot[black,mark=none] table[x=t, y=output2] {convergence_DRL.dat};
		\legend{Proposed DRL algorithm,The DRL algorithm [13],The RL algorithm [12]}	
		\end{axis}
		\end{tikzpicture}	
	}
	\caption{Comparisons of convergence performance.} \label{}

\end{figure}
\subsection{Simulation settings}
 We consider a scenario with a maximum of 100 UEs and 10 MEC servers distributed over a 500m x 500m area. The configurations for simulation parameters are listed in TABLE I. For the proposed DRL algorithm, motivated by [15], the proposed DNN structure includes an input layer, three hidden layers (64, 32 and 32 neurons), and an output layer. The discount factor $\gamma$ equals 0.85; the replay memory capacity and training batch size are set to $10^5$ and 128, respectively. The learning rate $\alpha$ is set to 0.01 to ensure stable and good performance for the algorithm. Further, we prepared 100,000 storing history samples to train the deep network for user selection and resource allocation decisions. Here, we employ ReLU as the activation function in the hidden layers, while the sigmoid activation function is utilized in the output layer to relax the decision variables. We used the AdamOptimizer to optimize the loss function. All simulations were implemented in Python 3.6 with TensorFlow 2.0 [15] on a computer with an Intel Core i7 4.7GHz CPU and 128 GB memory. 
 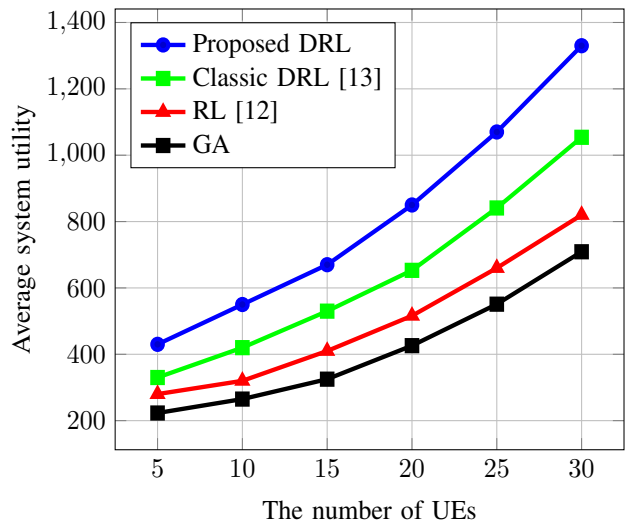
\begin{figure}
 	\centering
 	\resizebox {8.5cm} {7cm} {
 		\begin{tikzpicture}
 		\begin{axis}[legend pos=north west, legend cell align={left}, grid=major,
 		xlabel=The number of UEs,
 		ylabel= Average system utility, every axis plot/.append style={ultra thick}]
 		\addplot coordinates {
 			(5,430)
 			(10,550)
 			(15,670)
 			(20,850)
 			(25,1070)
 			(30,1330)

 		};
 		\addplot [mark=square*, green] coordinates {
 			(5,330)
 			(10,420)
 			(15,530)
 			(20,653)
 			(25,841)
 			(30,1104-50)
 			
 		};
 		\addplot [mark =triangle*, red] coordinates {
 			(5,280)
 			(10,320)
 			(15,410)
 			(20,516)
 			(25,660)
 			(30,820)
 			
 		};
 		\addplot [mark =square*, black] coordinates {
 			(5,223)
 			(10,265)
 			(15,325)
 			(20,426)
 			(25,551)
 			(30,709)	
 			
 		};
 		\legend{Proposed DRL, Classic DRL [13], RL [12], GA}	
 		\end{axis}
 		\end{tikzpicture}	
 	}
 	\caption{Average system utility versus numbers of UEs.} \label{}
 	\vspace{-0.15in}
 \end{figure}
\begin{table}
	\caption{Simulation parameters.}
	\label{table}
	\scriptsize
	\centering
	\captionsetup{font=scriptsize}
	\setlength{\tabcolsep}{5pt}
	\begin{tabular}{p{5cm}|p{2.5cm}}
		\hline
		\textbf{Parameter}& 
		\textbf{Value}
		\\
		\hline
		Number of UEs $U$& [5-30]
		\\
		Number of MEC servers $M$& 10
		\\
		Task's offloading data sizes $d_u$& [1-5] MB [1]
		\\
		Numbers of CPU cyles $\xi_u$& [0.6-1.6] Gcyles [11]
		\\
		The MEC server's computing capability $f^m_u$ & 5 GHz [14]
		\\
		The unit edge service price $\mu$ & 0.15 token/Gcyles [6]
		\\
		The hash power allocated to MEC servers $p^m$& [20-100] MHash/s [12]
		\\
		Numbers of transactions per block (block size) $b_m$ & [1-10]KB [16]
		\\
		The reward of the first miner which achieves consensus $\mathcal{R}$ & 30 tokens [16]
		\\
		\hline
	\end{tabular}
	\label{tab1}
\end{table}

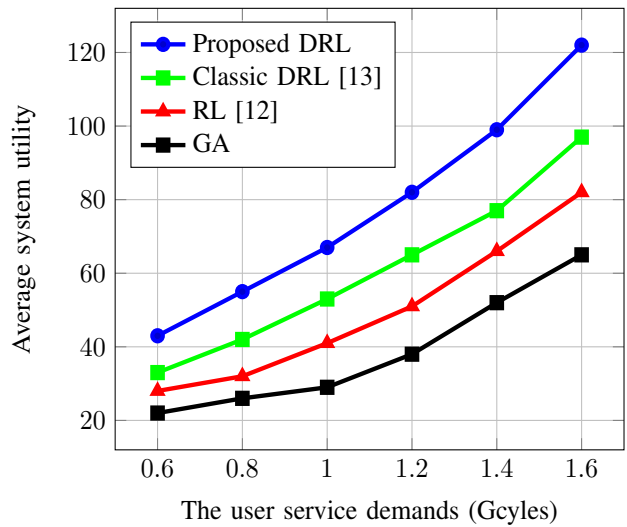
\begin{figure}
	\centering
	\resizebox {8.5cm} {7cm} {
		\begin{tikzpicture}
		\begin{axis}[legend pos=north west, legend cell align={left}, grid=major,
		xlabel=The user service demands (Gcyles),
		ylabel= Average system utility, every axis plot/.append style={ultra thick}]
		\addplot coordinates {
			(0.6,43)
			(0.8,55)
			(1.0,67)
			(1.2,82)
			(1.4,99)
			(1.6,122)
		};
		\addplot [mark=square*, green] coordinates {
			(0.6,33)
			(0.8,42)
			(1.0,53)
			(1.2,65)
			(1.4,77)
			(1.6,97)
			
		};
		\addplot [mark =triangle*, red] coordinates {
			(0.6,28)
			(0.8,32)
			(1.0,41)
			(1.2,51)
			(1.4,66)
			(1.6,82)
			
		};
		\addplot [mark =square*, black] coordinates {
			(0.6,22)
			(0.8,26)
			(1.0,29)
			(1.2,38)
			(1.4,52)
			(1.6,65)	
			
		};
		\legend{Proposed DRL, Classic DRL [13], RL [12], GA}	
		\end{axis}
		\end{tikzpicture}	
	}
	\caption{Average system utility versus user service demands.} \label{}
	\vspace{-0.15in}
\end{figure}

We compared our scheme (Proposed DRL) with the RL scheme (using a basic Q-learning algorithm) [12] and the classic DRL [13] (using a standard DQN algorithm). Moreover, genetic algorithm (GA), a common approach to solve nonlinear integer programming problems, is also considered. With the help of double DQN associated with appropriate parameter settings, the proposed DRL scheme should achieve the best performance, while the merit of other schemes can be shown via simulations.

\subsection{Simulation Results}
We verify the system utility performance of our scheme and compare it with other baseline approaches. Then we also analyze the optimization performance among schemes.
\subsubsection{Convergence performance}
We first evaluate the convergence performance of learning algorithms, including the proposed DRL, classic DRL and RL schemes for a blockchain –based MEC system with 10 MEC servers and 20 UEs. Fig. 4 shows the learning curves for total system utility obtained by running the learning algorithms over 2000 timeslots. The result shows that the total system utility is small at the beginning of the learning process for all schemes. However, as the number of timeslot increases, the total system utility of the proposed DRL scheme increases rapidly and converges after only 250 timeslots with high stability. An explanation is that the proposed DRL scheme uses Double DQN that reduces the overestimation of Q values and, as a result, accelerates the learning process. Moreover, the use of two separate value functions helps to better evaluate trained data for better algorithm performance (i.e. system utility).  Meanwhile, the classic DRL algorithm using a traditional deep D-learning with no further algorithm improvements converges slower (about 400 timeslots) with a lower system utility. The RL algorithm has the lowest convergence performance, which is because that the RL scheme with Q-learning always chooses system actions in a greedy manner, leading to slow action value evaluation. The comparison results verifies the superior performance of our DRL scheme, compared to its counterparts. 

\subsubsection{System utility performance}
We investigated the performances of four algorithms in various metrics as shown in Figs. 5,6. In specific, in Fig. 5, we consider a MEC network with 10 MEC servers and a varying number of UEs from 5 to 30, and measure the average system utility. It is observed that more UEs lead to a higher system utility due to the increase of system revenue and blockchain reward (i.e. mining profits). The GA scheme achieves the lowest utility in all cases of UE numbers. The main reason behind that that the higher number of UEs would result in a larger search space, which makes the GA scheme actually inefficient in finding an optimal optimization value and thus limits the performance of the algorithm. The RL shows a better performance in terms of higher system utility, compared with the GA scheme. Meanwhile, for the two DRL approach, when the number of UEs $U$ is small, fewer training steps are needed to meet system performance requirements, which leads to a similar system utility. However, when $U$ increases, the proposed DRL scheme with the help of double DQN learning is the best with a larger performance gap among all learning schemes and is much better than the GA scheme. 

Moreover, the Fig. 6 depicts the performance curves of average system utility with respect to the user service demands. In this case, we consider a scenario with 10 MEC servers and 20 UEs, each UE has a service demand varying from 0.6 to 1.6 Gcyles. It can be seen that with higher service demands, the incomes of the MEC system increase accordingly (as shown in equation (1)), which leads to an increase of the overall system utility although it also results in higher computation latency. Again, among all schemes, the proposed DRL scheme with a double DQN design exhibits the best performance with the largest average system utility, followed by the classic DRL scheme.

Next, we also evaluate the performance of the MEC system by our design with the joint optimization of user selection and resource allocation (i.e. power hash) versus the varying block sizes. In particular, we highlight the merit of the proposed scheme by comparing with the classic DRL scheme in terms to three scenarios, namely the full design with taking both user selection and resource allocation into consideration, the scheme without resource allocation and the scheme without user selection. As depicted in Fig. 7, our design with the proposed DRL scheme associated with a joint optimization of user selection and resource allocation yields the highest performance in terms of system utility when varying block sizes. The key reason is that when the larger block size means that the MEC servers obtain more rewards via mining (based on equation 6), which thus leads to a higher system utility. Although the other schemes with the optimization of only resource allocation or only user selection achieve lower system utilities, our DRL scheme is much better than the traditional DRL scheme, which shows the efficiency of our algorithm. 

\subsubsection{Optimization performance}
Next, we evaluate the performance of four algorithms through the average total edge computation latency-ECL (in Equation 2, in second) for a MEC network of 10 MEC servers and varying numbers of UEs, and average algorithm running time-ART (CPU computation time, in second) as presented in TABLE II. Note that the ECL results are averaged over the runs of 100,000 history samples for offloading action generation and DNN training. Overall, the ECL and ART increase rapidly in all schemes when the number of UEs increases due to a larger system state space and higher computing complexity that needs to be solved. Compared to the learning schemes, the GA algorithm consumes great computation resources, while its ECL is also high in all cases. In particular, the proposed DRL scheme achieves better ECL performances but requires less execution latency than its counterparts, showing the minimal computational complexity of our algorithm. 
\begin{figure}
	\centering
	\resizebox {8.5cm} {7cm} {
		\begin{tikzpicture}
		\begin{axis}[legend pos=north west, legend style={font=\tiny},legend cell align={left}, grid=major,
		xlabel= The block size (KB),
		ylabel=Average system utility, label style={font=\small},
		tick label style={font=\tiny} ]
		\addplot [mark=square*, red] coordinates {
			(1,2000)
			(2,2200)
			(3,2400)
			(4,2600)
			(5,2800)
			(6,2950)	
			(7,3100)	
			(8,3200)	
			(9,3300)	
			(10,3400)	
		};
		\addplot [mark=square*, orange]coordinates {
			(1,1800)
			(2,2000)
			(3,2200)
			(4,2400)
			(5,2600)
			(6,2700)	
			(7,2800)	
			(8,2900)	
			(9,3050)	
			(10,3100)	
		};
		\addplot [mark = triangle*, blue] coordinates {
			(1,1200)
			(2,1400)
			(3,1600)
			(4,1800)
			(5,1950)
			(6,2100)	
			(7,2200)	
			(8,2300)	
			(9,2400)	
			(10,2500)	
		};
		\addplot [mark = triangle*, green] coordinates {
			(1,1100)
			(2,1300)
			(3,1500)
			(4,1700)
			(5,1850)
			(6,2000)	
			(7,2100)	
			(8,2200)	
			(9,2300)	
			(10,2400)	
		};
		\addplot [mark = diamond*, brown] coordinates {
			(1,850)
			(2,1050)
			(3,1250)
			(4,1450)
			(5,1600)
			(6,1750)	
			(7,1850)	
			(8,1950)	
			(9,2080)	
			(10,2150)	
		};
		\addplot [mark = diamond*, black] coordinates {
			(1,800)
			(2,1000)
			(3,1200)
			(4,1400)
			(5,1550)
			(6,1700)	
			(7,1800)	
			(8,1900)	
			(9,2030)	
			(10,2100)	
		};
		\legend{The proposed DRL scheme, The classic DRL scheme [13],Proposed DRL scheme w/o resource allocation, Classic DRL scheme [13] w/o resource allocation, Proposed DRL algorithm w/o user selection,Classic DRL scheme [13] w/o user selection }	
		\end{axis}
		\end{tikzpicture}	
	}
	\caption{Average system utility versus block size $b_m$.} \label{}
	\vspace{-0.15in}
\end{figure}
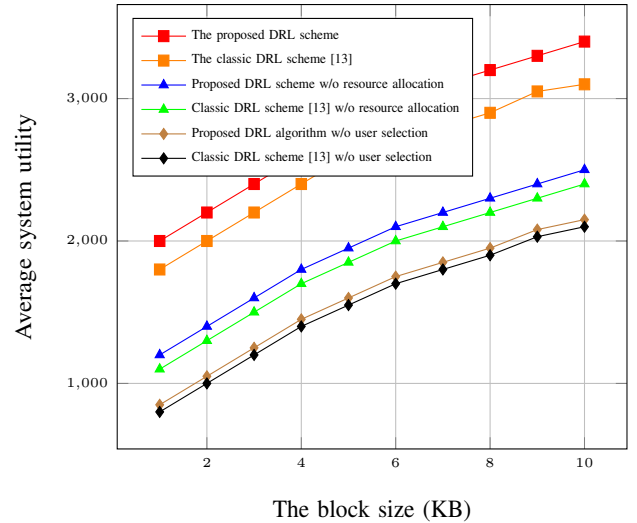
\begin{table}
	\scriptsize
	\centering
	\captionsetup{font=scriptsize}
	\caption{{Comparison results of algorithm performances.}}
	\label{table}
	
	\begin{tabular}{|c||c c | c c| c c|}
		\hline
		\multirow{2}{*}{\textbf{Schemes}} &
		
		\multicolumn{2}{c|}{\textit{N =10}} & \multicolumn{2}{c|}{\textit{N=20}} & \multicolumn{2}{c|}{\textit{N=30}} \\ 
		& {ECL} & {ART}  & {ECL} & {ART}& {ECL} & {ART}  \\
		\hline
		GA & 63.03 & 12804& 89.05 & 18048& 145.97 & 26889\\
		RL [12] & 61.54 & 29.12& 84.21 & 131.02& 119.34 & 538.33\\
		Classic DRL [13] & 59.67 & 26.09& 80.90 & 127.71& 110.38 & 478.19\\
		Proposed DRL& \textbf{53.42} & \textbf{21.54}& \textbf{74.09}& \textbf{122.38}& \textbf{99.05} & \textbf{445.63} \\
		\hline
		
	\end{tabular}
	\vspace{-0.1in}
\end{table}

\subsection{Smart contract performance}
We further investigate the feasibility of our smart contract, an important service in our BaaS model. Here, a trading contract is used to perform resource trading for offloading as mentioned in Section III.A. The contract mainly provides two main functions as follows:

- \textbf{CreationTrade():} This function includes user information $(W_u, Id_u,PK_u)$ and user $u$'s service information (including the service demand $\xi_u$ and service price $\mu$). It initializes the trading contract and runs it on blockchain for a ready execution of new trading requests.

- \textbf{Trading():} Based on user service demand $\xi_u$, the edge service provider (MEC server) calculates the service cost that the user $u$ needs to pay. During \textit{Trading()}, the user must complete their payment by sending a certain coin from their wallet ($W_u$) to the MEC server's wallet address.  

\begin{table}
	\scriptsize
	\centering
	\captionsetup{font=scriptsize}
	\caption{{Smart contract cost test.}}
	\label{table}
	
	\begin{tabular}{|c||c c c|}
		\hline
		\textbf{Contract functions}  & \textbf{Gas used} & \textbf{Actual cost (ether)} & \textbf{USD} \\
		\hline
		CreationTrade() &	170948 &	0.0034 &0.6630\\
		Trading() &	3904827	 &0.07809&	15.2275\\
		\textbf{Total} & \textbf{4075775} & \textbf{0.0815}& \textbf{15.8955}\\
		\hline
	\end{tabular}
	\vspace{-0.25in}
\end{table}

To evaluate the performance of the trading contract, we implemented a real-world experiment with a simplified blockchain-based MEC system according to our recent works [10], [15]. Here, we employed Ethereum, a popular blockchain platform for edge computing to create a blockchain network where there are five smart phones acting as UEs and one MEC server. The trading contract was written by Solidity programming language with two functions as mentioned above. We focus on evaluating the operation costs of contract functions as listed in TABLE III. 

Here, the cost is calculated in a gas unit and then converted into ether (cost unit of Etherum blockchain) and US dollars by using an exchange rate of 1 Gas $\approx$ 0.00000002 Ether and 1 Ether $\approx$ \$195, according to [19]. We consider a realistic scenario that some  users join the blockchain network for resource trading with the MEC server so that they can offload data, and thus $CreationTrade()$ and $Trading()$ functions need to be executed. All of these contract executions incur operation costs and the users need to pay for their service usage. From TABLE III, the amount of gas used for trading services is 4075775 gas (15.8955 USD, $\approx$ 3.1791 USD per user). Clearly, the financial cost for using our contract is low, which demonstrates the practicality of the proposed contract-based trading scheme.
\section{Conclusions}
In this paper, we have considered a multi-access edge computing (MEC) network with blockchain for multi-users. Our key focus is on the optimization of the performance of the MEC system by maximizing edge service revenue and blockchain mining reward and minimizing the service computation latency with respect to constraints of user service demands and hash power resources. Particularly, we leverage a smart contract to support transparent resource trading for computation offloading, while a reputation mechanism is designed to accelerate the mining process. We have formulated the user selection and resource allocation as a joint optimization problem. Then, a novel deep reinforcement learning (DRL) scheme using a double DQN algorithm is developed to solve the proposed problem. We have validated the performances of our design in various parameter settings, showing the significant improvements to the proposed scheme, compared with other baseline approaches in terms of much higher system utility. System evaluations also proved that the proposed algorithm achieved minimal computational complexity, and the operation cost of the smart trading contract was low, showing the efficiency of the proposed scheme.

\end{document}